\documentclass[12pt]{article}

\usepackage{fullpage}
\usepackage{setspace}
\usepackage{verbatim}
\usepackage{multirow}
\usepackage[round]{natbib}

  \usepackage{graphicx}
\title{Accounting for the influence of aquifer heterogeneity on spatial propagation of pumping drawdown}
\author{Dylan R. Harp (dharp@lanl.gov) and Velimir V. Vesselinov (vvv@lanl.gov)\\Earth and Environment Science Division\\Los Alamos National Laboratory}

\begin{document}
\maketitle

\begin{abstract}
It has been previously observed that during a pumping test in heterogeneous media, drawdown data from different time periods collected at a single location produce different estimates of aquifer properties and that Theis type-curve inferences are more variable than late-time Cooper-Jacob inferences. In order to obtain estimates of aquifer properties from highly transient drawdown data using the Theis solution, it is necessary to account for this behavior. We present an approach that utilizes an exponential functional form to represent Theis parameter behavior resulting from the spatial propagation of a cone of depression. This approach allows the use of transient data consisting of early-time drawdown data to obtain late-time convergent Theis parameters consistent with Cooper-Jacob method inferences. We demonstrate the approach on a multi-year dataset consisting of multi-well transient water-level observations due to transient multi-well water-supply pumping. Based on previous research, transmissivities associated with each of the pumping wells are required to converge to a single value, while storativities are allowed to converge to distinct values.\end{abstract}


%
%

%


%
%

\section{Introduction}

Aquifer property inferences obtained using the Theis type-curve method \citep{Jacob40} (Theis method) and the Cooper-Jacob straight-line approximation method \citep{Cooper46} (Cooper-Jacob method) at a given location have been observed to differ \citep{Ramey82,Butler90}. Theoretical investigations by \cite{Dagan82} utilizing a perturbation expansion approach on idealized scenarios demonstrate that effective hydraulic conductivity (transmissivity in 2D) decreases from the arithmetic mean conductivity to a convergent value over time. More recent numerical and field investigations demonstrate that Theis solution parameters \citep{Theis35} estimated at a location at various times during a pumping test have been observed to decrease at early times converging to stable values at late-times \citep{Wu05,Straface07}. \cite{Butler90} contributes this characteristic of Theis solution parameters to the fact that at early times, while the cone of depression is approaching the observation location, the drawdown is affected by many factors, such as: skin effects; well loses; and aquifer heterogeneities encountered by the cone of depression, complicating the estimation of stable parameters. However, at late times when quasi-steady state conditions have developed (i.e.\ when pressure gradients have reached steady state but pressures remain transient), the stable parameter estimates are consistent with aquifer property inferences that would be obtained using the Cooper-Jacob method. This implies that the late-time parameter estimates provide interpreted aquifer properties (as defined by \cite{Sanchez06}) representative of the support scale defined by the distance between the pumping and monitoring wells \citep{Neuman90,Neuman03c}.

Obtaining variable model parameter inferences indicate the inadequacy of a model to represent a system, as parameters are intended to represent invariant intrinsic properties of the system. The limitations of applying the Theis solution to model typical pumping tests is not a matter of debate, as its inadequacies are readily apparent by the assumptions required in its derivation \citep{Theis35} (e.g.\ fully penetrating well, infinite lateral extents, homogeneous properties, unperturbed initial conditions, confined aquifer). Recognizing these limitations, the question becomes whether or not the model can be useful. We agree with previous researchers that the Theis solution is useful for obtaining aquifer property inferences that characterize the groundwater transport if late-time drawdown data is used consistent with the Cooper-Jacob method \citep{Butler90,Meier98,Sanchez99,Knudby06,Trinchero08}. As noted by \cite{Butler90} in reference to the use of the Cooper-Jacob method, the advantage of drawing inferences from late-time drawdown data is that the estimated parameters will be independent of the numerous early-time effects that can influence the drawdown at the initial stages of expansion of the cone of depression. 

Obtaining late-time pumping data at quasi-steady state is not always possible, however, as it may not be feasible to continue a pumping test for a sufficient duration to allow quasi-steady state conditions to develop. Or, in cases where an existing water-supply and water-level elevation dataset is available from a municipal water supply network, quasi-steady state may not be reached due to a high frequency of cycling multiple supply wells on and off to: meet shifting demand; to take advantage of lower cost off-peak electrical rates; and perform well maintenance and/or repair. In this paper, we present an approach that allows convergent parameters to be obtained from transient drawdown data by accounting for the behavior of Theis parameters at early times. 

The proposed approach is demonstrated on a long-term highly-transient drawdown record collected at the Los Alamos National Laboratory (LANL) site where the water-level transients result from multi-well municipal water-supply pumping. The pumping regimes are highly transient, cycling diurnally and seasonally, including variations due to maintenance, repair, and shifting supply loads within the network. As a result, the drawdown at monitoring wells within the network do not fully attain quasi-steady state as new pressure influences begin to propagate through the aquifer as the pumping wells cycle on and off \citep{Harp10}. The use of a long-term dataset containing multiple pressure influence cycles has certain advantages, such as: reduction of measurement errors; improved characterization of the hydraulic response allowing the refinement of hydrogeologic inferences; and the lack of the expense and coordination of a conventional pumping test. We demonstrate the inference of aquifer properties from this dataset by considering the transient early-time behavior of Theis solution parameters.

As the approach presented here utilizes observations, numerical experiments, and analytical investigations of many previous researchers \citep{Dagan82,Ramey82,Butler90,Meier98,Sanchez99,Wu05,Knudby06,Straface07,Trinchero08}, a review of these bodies of research will be presented in the background section. The proposed approach for accounting aquifer heterogeneity on Theis parameters will be presented in the methodology section. The approach will be demonstrated on the LANL dataset in the results section.

\section{Background} 

It has been recognized that aquifer property inferences based on the Theis method and Cooper-Jacob method differ \citep{Ramey82}. This is due to the fact that the inference methods emphasize properties in different regions of the aquifer. The Theis method considers the entire drawdown curve, often leading to an emphasis on the interval of greatest curvature located during early times. As indicated by \cite{Butler90}, drawdown at early times can be affected by many factors, including local heterogeneities near the pumping well and well skin and pumping storage affects, creating greater variability in Theis method inferences. The Cooper-Jacob method ignores early times, providing information on the properties of the aquifer within a ring formed by the outward moving front of the cone of depression during the time interval under consideration. At late time, when the Cooper-Jacob approximation is valid, the region included in this ring can be large. \cite{Butler90} demonstrates that the difference between inferences obtained from the Theis and Cooper-Jacob methods depend on the level of aquifer heterogeneity and the distance between the pumping well and the observation location. The inferences become more similar as the level of heterogeneity decreases and the distance increases.  

\cite{Meier98} explore the use of the Cooper-Jacob approximation to infer effective transmissivity ($T_{eff}$) from the estimated transmissivity parameter $\hat{T}$ and provide indications of hydraulic connectivity by evaluating the estimated storativity parameter $\hat{S}$ in heterogeneous aquifers. Consistent with theoretical findings of \cite{Butler90}, \cite{Meier98} present cases where field data demonstrate that although small-scale (point) estimates of transmissivity $T$ are highly variable, values of $\hat{T}$ obtained from the Cooper-Jacob method are relatively constant.  Furthermore, \cite{Meier98} demonstrate that corresponding values of $\hat{S}$ are typically highly variable, even though the storativity in the field is believed to be relatively constant. \cite{Meier98} investigate this phenomena performing numerical experiments with heterogeneous transmissivity fields and homogeneous storativity fields, producing similar values for $\hat{T}$ and variable values for $\hat{S}$ consistent with field cases. 

The reason for this paradoxical result can be explained by examining the equation for estimating $T$ from the Cooper-Jacob method; $\hat{T}=2.3Q/4\pi I$, where $Q$ is a constant pumping rate and $I$ is the slope of the late-time drawdown with respect to the (base 10) log of time (i.e.\ $I=(s_2-s_1)/(log\ t_2-log\ t_1)$, where $s$ is drawdown and $t$ is time). This equation demonstrates that $\hat{T}$ is dependent on the rate of drawdown decline, which is dependent on the choice of $t_2$ and $t_1$. Considering only the late-time drawdown where the data approximate a straight line with respect to log time, in accordance with the Cooper-Jacob method, means that the $\hat{T}$ will approximate $T_{eff}$ described by the rate of drawdown after the drawdown cone of depression has passed the monitoring well. Storativity estimates using the Cooper-Jacob method (defined as $\hat{S}=2.25Tt_0/r^2$, where $r$ is the distance from the pumping well to the observation point and $t_0$ represents the time-axis intercept of a line drawn through the late-time drawdown), on the other hand, are dependent on the variability of $T$ between the pumping well and the front of the cone of depression. Although the heterogeneity between the pumping and monitoring well does not affect the slope of the late-time drawdown used to determine $\hat{T}$, it can affect $\hat{S}$ as the time-axis intercept ($t_0$) is dependent on the arrival of the cone of depression at the monitoring well. If a region of high $T$ connects the monitoring well and the pumping well, the value of $t_0$ will be relatively small and vice-versa. As noted by \cite{Sanchez99}, the Cooper-Jacob method interprets an early/late arrival of a drawdown cone of depression as low/high storativity. This explains the high variability of $\hat{S}$ in the presence of $T$ heterogeneity between the pumping well and the monitoring well, even in cases where $S$ is known to be constant. 

Research by \cite{Meier98} demonstrate from a numerical analysis that $\hat{T}$ estimated from a simulated pumping test (radial flow) is close to $T_{eff}$ for parallel flow for an area of influence for multilognormal stationary (geostatistically homogeneous) $T$ fields (the $S$ field is assumed uniform in all cases). While \cite{Meier98} also demonstrated that this can be true for nonmultigaussian fields, this is not necessarily true in general \citep{Sanchez96}. Similar to findings by \cite{Butler88}, who demonstrated that $\hat{S}$ depends on transmissivities between the pumping well and the front of the cone of depression, \cite{Meier98} find that $\hat{S}$ depends on transmissivities between and nearby the well and the observation point.

\cite{Sanchez99} verify these conclusions using an analytical approximation to the groundwater flow equation. They demonstrate analytically that $\hat{T}$ is independent of spatial location. They also demonstrate that storativity estimates will provide an indication of the local deviations of $T$ from its large-scale geometric mean (denoted as $T_G$) representing the equivalent geostatistically homogeneous $T$ field. If $T$ in a specific location is less than $T_G$, $\hat{S}$ will be larger than the true value of $S$ and vice-versa. They also show that the geometric mean of $\hat{S}$ values is an unbiased estimator of $S$.

\cite{Knudby06} demonstrate that Cooper-Jacob estimates of diffusivity ($\hat{D}=\hat{T}/\hat{S}$) correlate well with indicators of flow and transport connectivity. \cite{Trinchero08} demonstrate that estimated effective porosity (a transport connectivity indicator) depends on a weighted function of actual transmissivity and the interpreted Cooper-Jacob storativity along the flow line.


In contrast to \cite{Meier98}, \cite{Sanchez99}, \cite{Knudby06}, and \cite{Trinchero08}, \cite{Wu05} explore the effect of the homogeneity assumption of the Theis solution on parameter estimates for the entire drawdown curve (including early and late time data) for cases with heterogeneous $T$ and $S$ fields. Conceptualizing $T$ and $S$ as spatial stochastic processes in the equation of flow, \cite{Wu05} derive the mean flow equation of a heterogeneous confined aquifer as

\begin{equation}
	T_{eff} \nabla^2 \langle h \rangle = S_{eff} \frac{\partial \langle h \rangle }{\partial t}
	\label{eq:stoch_theis}
\end{equation}

\noindent where angle brackets indicate ensemble mean, $t$ is time, $T_{eff}$ is the effective transmissivity defined as 

\begin{equation}
	T_{eff} = \overline{T} + \frac{\langle T'\nabla h'\rangle}{ \nabla \langle h \rangle},
\end{equation}

\noindent and $S_{eff}$  is the effective storativity defined as

\begin{equation}
	S_{eff} = \overline{S} + \frac{\left \langle S' \frac{\partial h'}{\partial t} \right \rangle}{ \frac{\partial \langle h \rangle}{\partial t} }.
\end{equation}

\noindent where the over bar and prime denote the spatial mean and perturbation of the variable, respectively. $T_{eff}$ and $S_{eff}$ are denoted as effective parameters as they will produce the ensemble mean head $\langle h \rangle$ as a convergent average for a set of realizations of heterogeneity based on the stochastic parameters $T=\overline{T} + T'$ and $S=\overline{S} + S'$. As indicated by \cite{Wu05}, in order for the ensemble mean head $\langle h \rangle$ to equal the spatially averaged head $\overline{h}$ of a single realization of heterogeneity, the field must contain an adequate sampling of the heterogeneity (i.e.\ the field must be ergodic). As traditional pumping tests typically estimate $T_{eff}$ and $S_{eff}$ based on one or a small number of point estimates of head, which will not equal the spatially averaged head in general, $\hat{T}$ and $\hat{S}$ will not provide estimates of effective properties in an ensemble sense in general.


\cite{Wu05} performed numerical experiments using synthetic aquifers with multilognormal heterogeneous $T$ and $S$ fields. They observe that at early time, $\hat{T}$ estimates at different locations are highly variable, while, similar to the findings of \cite{Meier98} and \cite{Sanchez99}, at large times (when the Cooper-Jacob approximation is valid) values of $\hat{T}$ converge to a value close to $T_G$ as the cone of depression expands for the multilognormal fields considered. As the considered transmissivity field is multilognormal, $T_G=T_{eff}$. In cases where the transmissivity is nonmultigaussian, the significance of $\hat{T}$ is less certain \citep{Sanchez99}, however, we assume that it is a good first estimate of $T_{eff}$. In the same analysis, \cite{Wu05} demonstrated that values of $\hat{S}$ do not converge to a single value, but stabilize relatively quickly to values predominantly dependent on the heterogeneity between the pumping well and the given monitoring location. 

Similarities to the numerical results of \cite{Wu05} can be seen in the analytical investigation by \cite{Dagan82}, who utilized a perturbation expansion approach to explore the temporal behavior of $K_{eff}=-\langle \mathbf{q} \rangle/ \nabla \langle h \rangle$, ($T_{eff}$ in 2D), where $\mathbf{q}$ is discharge. He derived an approximate relation describing the temporal behavior of $T_{eff}$ for the idealized case of sufficiently small transmissivity variance and average head gradient slowly varying spatially and temporally in a stationary random field as

\begin{equation}
	T_{eff}(t) = e^{\mu_Y} \left [ 1 + \frac{1}{2} \sigma_Y^2 b_2(t) \right ]
	\label{eq:tefft}
\end{equation}

\noindent where $\mu_Y$ is the mean and $\sigma_Y^2$ is the variance of $Y=ln(T)$ and $b_2(t)$ is a function describing the temporal dependency of $T_{eff}$ based on the aquifer heterogeneity in 2D, equal to unity for $t=0$ and tending to zero as $t \rightarrow \infty$. Recognizing that the limiting cases for equation~(\ref{eq:tefft}) are first-order approximations of the arithmetic mean transmissivity $T_A$ ($t=0$) and $T_{eff}$ ($t \rightarrow \infty$), $b_2(t)$ can be expressed as

\begin{equation}
	b_2(t) = \frac{T_{eff}(t) - T_{eff,c}}{T_A - T_{eff,c}}
	\label{eq:b_2}
\end{equation}

\noindent where $T_{eff,c}$ is the late-time convergent $T_{eff}$. Equation~(\ref{eq:b_2}) indicates that $b_2(t)$ describes the temporal decline of $T_{eff}$ from $T_A$ to $T_{eff,c}$.


In contrast to the four field cases discussed by \cite{Meier98} (i.e.\ Grimsel test site, Switzerland \citep{Frick92}; El Cabril site, Spain \citep{Minieres90}; Horkheimer Insel site, Germany \citep{Schad94}; and Columbus Air Force Base, U.S.A.\ \citep{Herweijer91}), \cite{Straface07} observe a lack of similar slope for drawdown vs log time  at late times from pumping tests near Montalto Uffugo Scalo, Italy, indicating that the Cooper-Jacob straight-line approximation for late-time drawdown will not be valid in all cases. Based on their analysis of these pumping tests, \cite{Straface07} question the validity of using traditional pumping tests to estimate meaningful hydrogeological parameters, but do suggest that these results can provide quick inexpensive first estimates. Furthermore, they suggest that these first estimates can be useful as starting parameters for a tomographic inversion of the same dataset.

\cite{Harp10} demonstrate an approach to identify and decompose the pressure influences at a monitoring location using the Theis solution. Their approach is demonstrated on the same dataset as in the current research. As the objective of the research in \cite{Harp10} is the decomposition of pressure influences, attempts are not made to account for early time behavior of the Theis solution parameters, and constant and distinct values are applied to pumping/monitoring well pairs. Therefore, the parameter estimates are not considered representative of the aquifer properties of the aquifer, but are interpreted parameters characterizing the hydraulic response at a monitoring location due to pumping a single well. These interpreted parameters are analogous to parameter estimates that would be obtained from a conventional pumping test analysis. 

The current research presents an approach to account for Theis parameter behavior to infer aquifer properties considering the extensive body of research presented above. While the current approach is demonstrated on the long-term dataset from the LANL site, providing the decomposition of pressure influences similar to the approach presented in \cite{Harp10}, the current approach could also be applied to a conventional pumping test to more appropriately account for the behavior of the Theis solution parameters. Furthermore, this could be particularly useful to obtain late time hydrogeologic inferences from conventional pumping tests that were not conducted for a sufficient length of time to establish quasi-steady state conditions.

\section{Methodology}

The Theis solution of the flow equation in homogeneous media ($T\nabla^2 h = S \partial h/\partial t$) is defined as

\begin{equation}
	s_p(t) = \frac{Q}{4 \pi T} W(u) = \frac{Q}{4 \pi T} W\left(\frac{r^2S}{4Tt}\right),
	\label{eq:theis}
\end{equation}

\noindent where $s_p(t)$ is the predicted pumping drawdown at time $t$ since the pumping commenced (i.e.\ $h(t) - h(0)$), $Q$ is the pumping rate, $T$ is the transmissivity, $W(u)$ is the negative exponential integral ($\int_u^{\infty} e^{-y}/y \,dy$) referred to as the well function, $u$ is a dimensionless variable, $r$ is radial distance from the pumping well, and $S$ is the storativity. Multiple pumping wells and variable rate pumping periods can be included in the Theis solution by employing the principle of superposition \cite[page 327]{Freeze79} as

\begin{equation}
	s_p(t) = \displaystyle \sum_{i = 1}^N \displaystyle \sum_{j = 1}^{M_i} \frac{Q_{i,j}-Q_{i,j-1}}{4 \pi T}W \left(\frac{r_i^2 S}{4 T (t-t_{Q_{i,j}})}\right),
	\label{eq:theis_super}
\end{equation}

\noindent where $N$ is the number of pumping wells (sources), $M_i$ is the number of pumping periods (i.e.\ the number of pumping rate changes) for pumping well $i$, $Q_{i,j}$ is the pumping rate of the $i$th well during the $j$th pumping period, and $t_{Q_{i,j}}$ is the time when the pumping rate changed at the $i$th well to the $j$th pumping period. The drawdown calculated by equation~(\ref{eq:theis_super}) represents the cumulative influence at a monitoring location of the $N$ pumping wells at distances $r_i,\ i=1,\ldots,N$ from the monitoring location. 

Equations~(\ref{eq:theis}) and (\ref{eq:theis_super}) are only valid under the assumption of homogeneity. If a system is homogeneous, then $T$ and $S$ in equations~(\ref{eq:theis}) and (\ref{eq:theis_super}) will be equivalent to $T_{eff}$ and $S_{eff}$, respectively. If the system is heterogeneous, this will only be true in an ensemble mean sense. In this case, the Theis solution can be expressed as

\begin{equation}
	\langle s_p \rangle(t) = \frac{Q}{4 \pi T_{eff}} W(u) = \frac{Q}{4 \pi T_{eff}} W\left(\frac{r^2S_{eff}}{4T_{eff}t}\right)
	\label{eq:stoch_s}
\end{equation}

\noindent which is the solution to equation~(\ref{eq:stoch_theis}) \citep{Wu05}, where $\langle s_p \rangle(t)$ is the ensemble mean drawdown due to pumping and heads have been converted to drawdown. Invoking superposition with equation~(\ref{eq:stoch_s}), an ensemble mean drawdown equation analogous to equation~(\ref{eq:theis_super}) can be expressed as

\begin{equation}
	\langle s_p \rangle(t) = \displaystyle \sum_{i = 1}^N \displaystyle \sum_{j = 1}^{M_i} \frac{Q_{i,j}-Q_{i,j-1}}{4 \pi T_{eff}}W \left(\frac{r_i^2 S_{eff}}{4 T_{eff}* (t-t_{Q_{i,j}})}\right).
	\label{eq:stoch_super}
\end{equation}

As water elevations recorded at monitoring wells in an aquifer system are merely point samples from a single realization of heterogeneity, and not ensemble mean values of multiple realizations or spatial averages of an ergodic field, application of equations~(\ref{eq:stoch_s}) and (\ref{eq:stoch_super}) are invalid for cross-hole interference tests. Recognizing this theoretical limitation of applying the Theis solution (or the Cooper-Jacob approximation) to data from heterogeneous aquifers to infer effective properties, researchers have investigated what information is contained in the hydrogeologic parameter estimates \citep{Meier98,Sanchez99,Wu05,Knudby06,Trinchero08}. We propose that although the Theis solution parameters will not provide precise representations of hydrogeological properties, the analytical framework of the Theis solution can provide initial estimates of the effective transmissivity and indications of connectivity.

Recognizing that a dataset containing drawdown outside of the Cooper-Jacob domain will require consideration of the behavior of parameter estimates at early times (as the front of the cone of depression is at short radial distance), we approximate the Theis solution, defining the estimated pumping drawdown $\hat{s}_p(t)$ as

\begin{equation}
	\hat{s}_p(t) = \displaystyle \sum_{i = 1}^N \displaystyle \sum_{j = 1}^{M_i} \frac{Q_{i,j}-Q_{ij-1}}{4 \pi \hat{T}_i}W \left(\frac{r_i^2 \hat{S}_i}{4 \hat{T}_i* (t-t_{Q_{i,j}})}\right),
	\label{eq:est_s}
\end{equation}

\noindent where $\hat{T}_i$ and $\hat{S}_i$ are time dependent functions describing the variation in interpreted transmissivities and storativities as the cone of depression propagates outward from the pumping well. In order to provide a general functional form with the intent to capture the temporal dependence of $\hat{T}$ and $\hat{S}$ for a broad range of heterogeneities and pumping well factors, $\hat{T}_i$ and $\hat{S}_i$ are defined using an exponential functional form as

\begin{equation}
	\hat{T}_i(t) = \hat{T}_{eff} e^{c_{T,i}/(t-t_{Q_{i,j}})}\ \ c_T \ge 0,
	\label{eq:exp_T}
\end{equation}

\noindent and

\begin{equation}
	\hat{S}_i(t) = \hat{S}_{a,i} e^{c_{S,i}/(t-t_{Q_{i,j}})},
	\label{eq:exp_S}
\end{equation}

\noindent where $\hat{T}_{eff}$ provides the late-time convergent estimate for $T_{eff}$, $\hat{S}_{a,i}$ provides late-time convergent indications of connectivity between the $i$th pumping well and the monitoring location \citep{Knudby06}, and $c_{T,i}$ and $c_{S,i}$ are constants describing the time dependent slope of the transmissivity and storativity parameters, respectively, associated with the $i$th pumping well. Since in most cases, the statistical nature of the heterogeneity will not be known with certainty, this ad hoc functional form is assumed reasonable. In idealized scenarios with known correlation structure, it may be possible to derive these relationships in an ensemble mean sense. For example, \cite{Dagan82} derives an analytical relationship describing the temporal behavior of $T_{eff}$ for an exponential autocorrelation.

Based on \cite{Dagan82} and \cite{Wu05}, we constrain $c_T\ge0$, indicating that $\hat{T}(t)$ values from early time portions of drawdown data are expected to be higher than late time convergent values. This may be explained by the early-time negative correlation between head and transmissivity \citep{Wu05} and/or the existence of high conductivity inter-well pathways as described by \cite{Herweijer96}. Other possible explanations for time-dependent hydrogeologic parameters are well-bore storage and leakage effects known to exist at the site \citep{McLin05,McLin06,McLin06a}. 

Substituting equations~(\ref{eq:exp_T}) and (\ref{eq:exp_S}) into equation~(\ref{eq:est_s}) produces

\begin{equation}
	\hat{s}_p(t) = \displaystyle \sum_{i = 1}^N \displaystyle \sum_{j = 1}^{M_i} \frac{Q_{i,j}-Q_{ij-1}}{4 \pi \hat{T}_{eff} e^{c_{T,i}/(t-t_{Q_{i,j}})}}W \left(\frac{r_i^2 \hat{S}_{a,i} e^{c_{S,i}/t-t_{Q_{i,j}}}}{4 \hat{T}_{eff} e^{c_{T,i}/(t-t_{Q_{i,j}})} * (t-t_{Q_{i,j}})}\right).
	\label{eq:est_s_t}
\end{equation}

In order to account for temporal trends identified in a previous study \citep{Harp10}, we include an additional drawdown term $\hat{s}_t(t)$ as

\begin{equation}
	\hat{s}_t(t) = (t-t_i) \times m
\end{equation}

\noindent where $t_i$ is the time at the beginning of the pumping record and $m$ is a constant defining the linear increase in drawdown not attributable to pumping. 

As the calibration targets in the model inversions presented here are water elevations as opposed to drawdowns, we define the predicted water elevation $\hat{h}(t)$ at time $t$ as

\begin{equation}
	\hat{h}(t) = \hat{h}_o - \hat{s}_p(t) - \hat{s}_t(t)
	\label{eq:head}
\end{equation}

\noindent where $\hat{h}_o=\hat{h}(0)$ and is defined as the initial predicted water elevation at the monitoring well. As defined by the Theis solution, $\hat{h}_o$ is the head at the time that a perturbation commences. As pumping of the regional aquifer began at the LANL site over 50 years ago, it is reasonable to assume that the influence of the earlier pumping has propagated through the system and/or dissipated. However, more recent pumping rate changes preceding pressure transient records at monitoring locations need to be considered. In order to account for residual effects of pumping prior to monitoring, simulations are started in advance of pressure transient records including earlier pumping records. Therefore, $\hat{h}_o$ is not a measured quantity, but predicted at the beginning of the simulations.

Model calibration is performed using a Levenberg-Marquardt approach \citep{Levenberg44,Marquardt63} where the objective function can be defined as

\begin{equation}
	\mathbf{\Phi} = \displaystyle \sum_{i=1}^m \displaystyle \sum_{j=1}^{n_i} [h_i(t_j)-\hat{h}_i(t_j)]^2
\end{equation}

\noindent where $m$ is the number of monitoring wells considered, $n_i$ is the number of head observations for the $i$th monitoring well, and $h_i(t_j)$ are the head observations for the $i$th monitoring well included as calibration targets where $j$ is an observation time index.

The simulation of the drawdowns is performed using the WELLS code (available upon request at http://www.ees.lanl.gov/staff/monty/), which implements equation~(\ref{eq:est_s_t}). The calibration is performed using \emph{PEST} \citep{Doherty04}.

\section{Site Data}
\label{sect:site}

The regional aquifer beneath the LANL site is a complex stratified hydrogeologic structure which includes unconfined zones (under phreatic conditions near the regional water table) and confined zones (deeper zones) \citep{Vesselinov04a, Vesselinov04b}. The three monitoring wells considered in this analysis are screened near the top of the aquifer with an average screen length of 11 meters. The water-supply wells partially penetrate the regional aquifer with screens that also begin near the top of the aquifer, but penetrate deeper with an average screen length of 464 meters. Nevertheless, field tests demonstrate that most of the groundwater supply is produced from a relatively narrow section of the regional aquifer that is about 200-300 m below the regional water table \citep{LANLPajIR08}. Implicit in the use of the Theis solution is the assumption that groundwater flow is confined and two-dimensional. We assume that this is a justifiable assumption here given the small magnitude of observed drawdowns (less than 1 m at the monitoring wells and less than 20 m at the water-supply wells) and the relative long distances between supply and monitoring wells compared to the effective aquifer thickness (about 200-300 m). It is believed that leakage and vertical flow may be significant factors within the aquifer system, however, lack of monitoring locations within aquifers and confining units and lack of sufficient information to identify the locations of aquifers and confining units hinders approaches that intend to account for these affects (e.g.\ \cite{Neuman72}).

\begin{figure}
	\begin{center}
		\includegraphics[width=12cm]{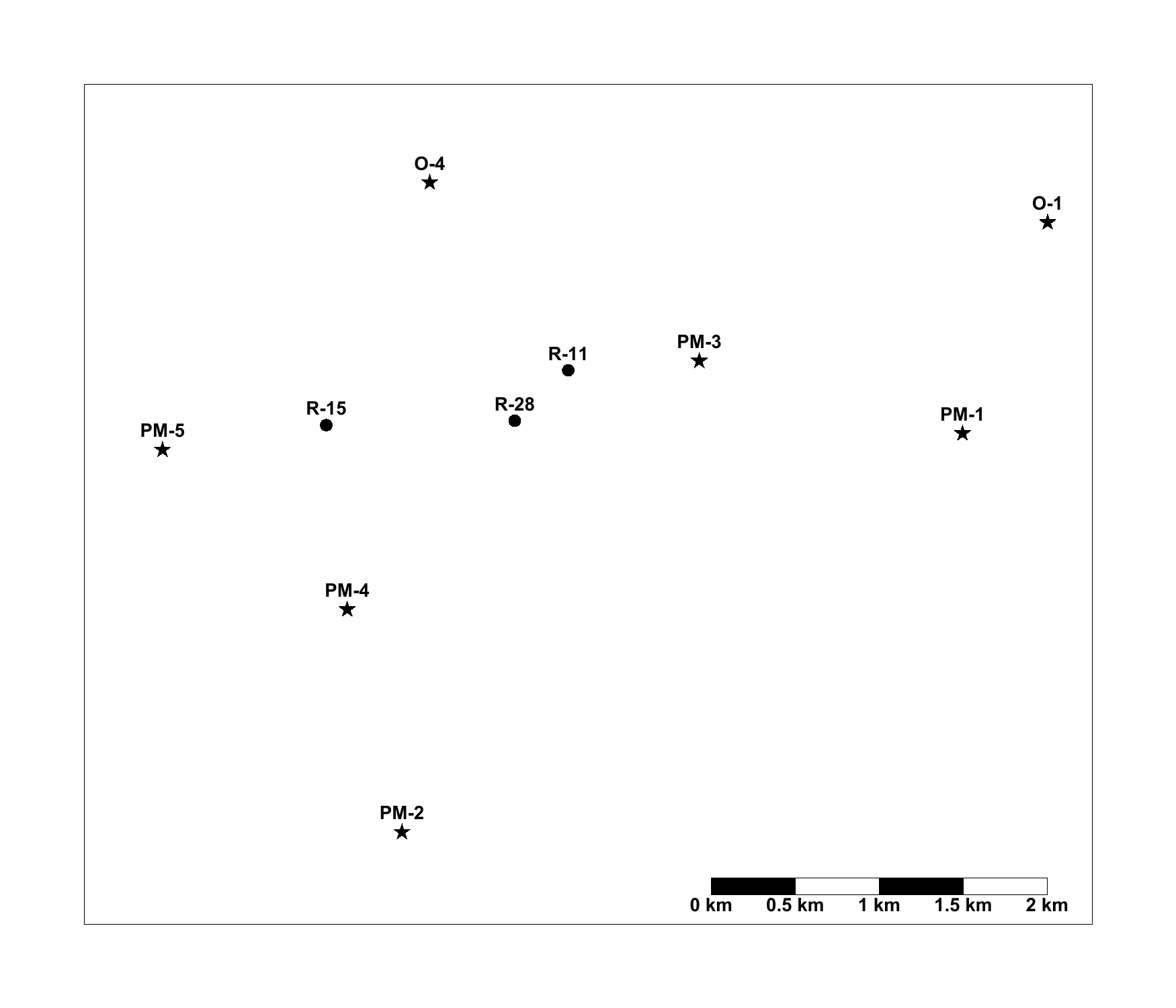}
		\caption{Map of monitoring wells (circles) and water-supply wells (stars) included in the analysis. Locations of newly completed and planned monitoring wells are indicated by open diamonds.}
		\label{fig:wells}
	\end{center}
\end{figure}

\begin{figure}
	\begin{center}
		\includegraphics[width=14cm]{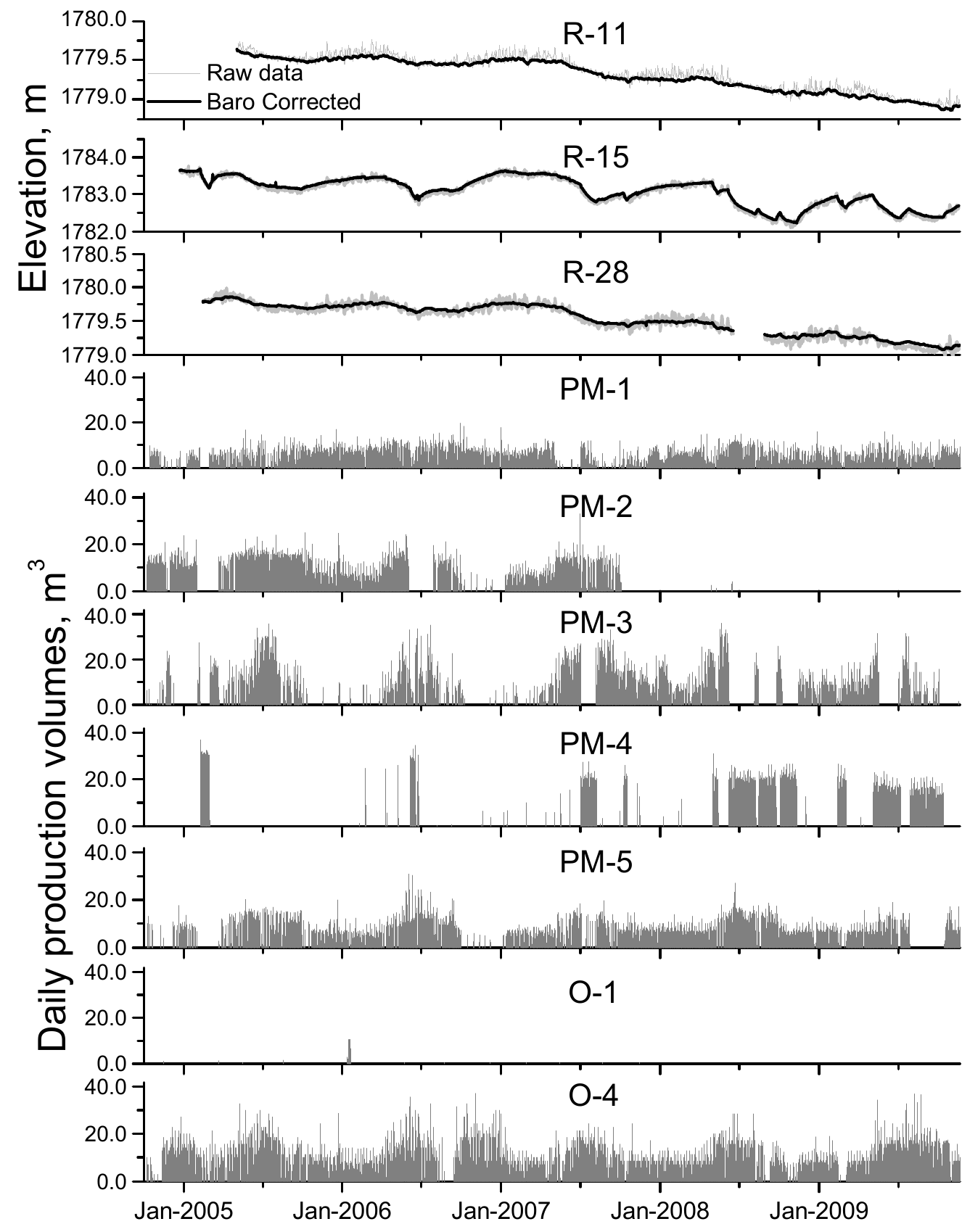}
		\caption{Water elevations at monitoring wells and production records for water-supply wells.}
		\label{fig:records}
	\end{center}
\end{figure}

Water-level fluctuations (pressure transients) are automatically monitored using pressure transducers. The pressure and water-supply pumping records considered here are collected from 3 monitoring wells (R-11, R-15 and R-28) and the 7 water-supply wells (PM-1, PM-2, PM-3, PM-4, PM-5, O-1, and O-4) located within the LANL site. Figure~\ref{fig:wells} displays a map of the relative location of the wells. Figure~\ref{fig:records} presents the pressure and production records for the monitoring wells and water-supply wells, respectively.

The water-level observation data considered here span approximately five years, commencing on or shortly after the date of installation of pressure transducers (May 4, 2005 for R-11; December 23, 2004 for R-15; February 14, 2005 for R-28), all terminating on November 20, 2009. The barometric pressure fluctuations are removed using constant coefficient methods using 100\% barometric efficiency for all monitoring wells \citep{LAUR4702}. Although the pressure transducers collect observations every 15 minutes, this dataset is reduced to single daily observations by using the earliest recorded measurement for each day. Some daily observations have been excluded due to equipment failure. Pumping records for all pumping wells begin on October 9, 2004 and terminate on December 31, 2009. The pumping record precedes the water-level calibration data to account for water-level transients due to pumping variations before the water-level data collection commenced. For additional information on the site and dataset, refer to \cite{Harp10}.  

In the applied computational framework, forward model run times for predicting water elevations at R-11, R-15, and R-28 for approximately four years (from October 8, 2004 to November 18, 2008) are each approximately 3 seconds on a 3.0 GHz Intel processor. Inversions initiated with uniform initial parameter values require approximately 600 model runs and, using a single processor, are performed for approximately 1 hour and 40 minutes.

\section{Results and Discussion}

Results of calibrations using temporally varying parameters and constant parameters are presented below. In the exponential case, a single value of $\hat{T}_{eff}$ is applied to all pumping/monitoring well pairs. Similarly, in the constant case, a single value of $\hat{T}$ is applied to all pumping/monitoring well pairs. Distinct values are allowed for $\hat{S}_a$ and $\hat{S}$ in the exponential and constant cases, respectively. Pumping influences (wells) that the calibration is unable to fingerprint at the monitoring well result in unrealistic parameter values that effectively eliminate the influence of the pumping well (i.e.\ high $\hat{T}$ and $\hat{S}$). As these parameter values are not physically meaningful beyond identifying a lack of influence from the associated pumping well, they are not presented below. Therefore, the omission of a pumping well below indicates a lack of identifiable influence at a monitoring well by the inversion.

\begin{figure}
	\begin{center}
		\includegraphics[width=14cm]{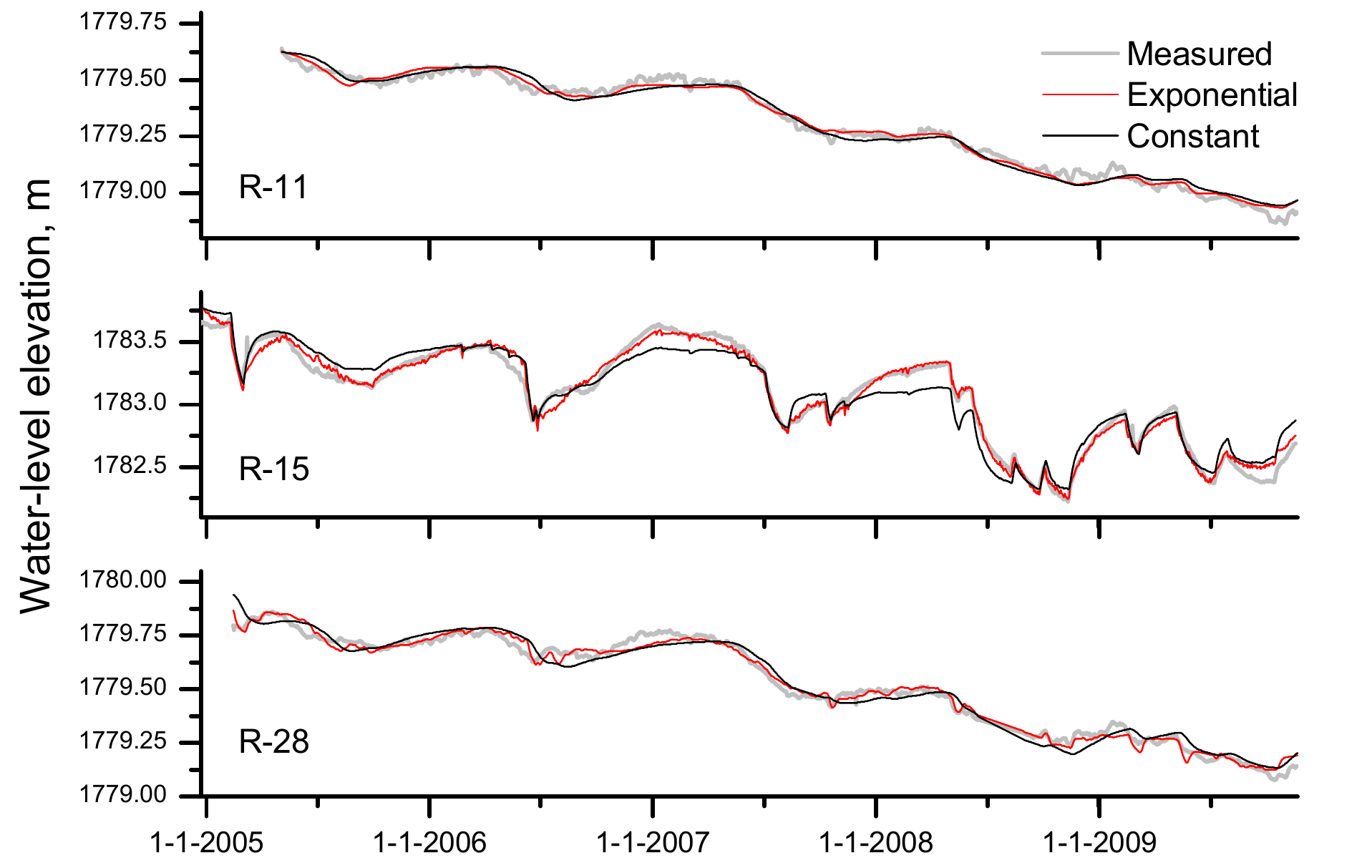}
		\caption{Calibrated heads for the exponential (red) and constant (black) cases. The observed heads are presented in gray.} 
		\label{fig:exp-calibs}
	\end{center}
\end{figure}

\begin{figure}
	\begin{center}
		\includegraphics[width=14cm]{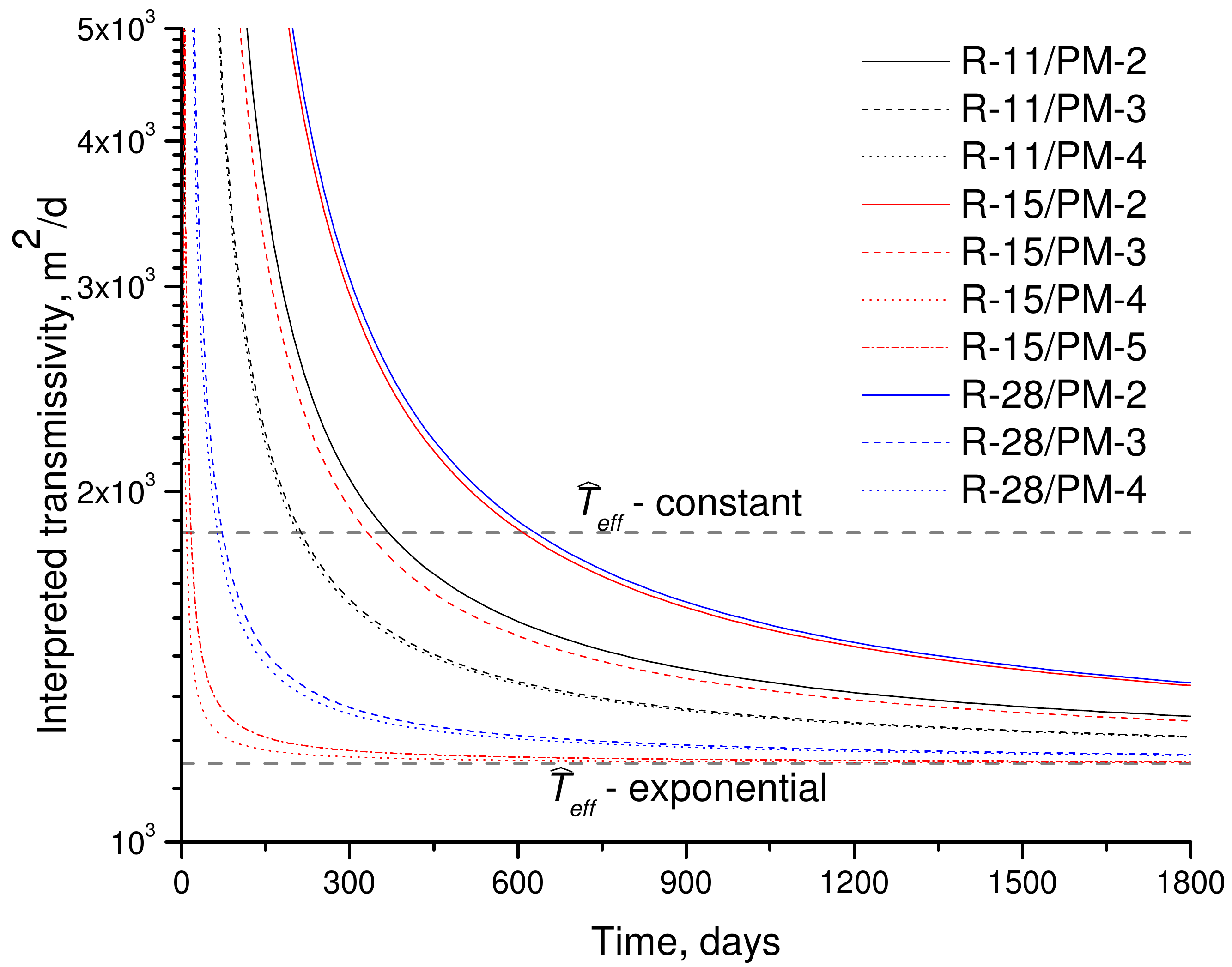}
		\caption{Estimated transmissivity functions for the exponential case. The convergent value of $\hat{T}_{eff}$ and $\hat{T}$ for the constant case are indicated.}
		\label{fig:all-trans}
	\end{center}
\end{figure}


\begin{figure}
	\begin{center}
		\includegraphics[width=14cm]{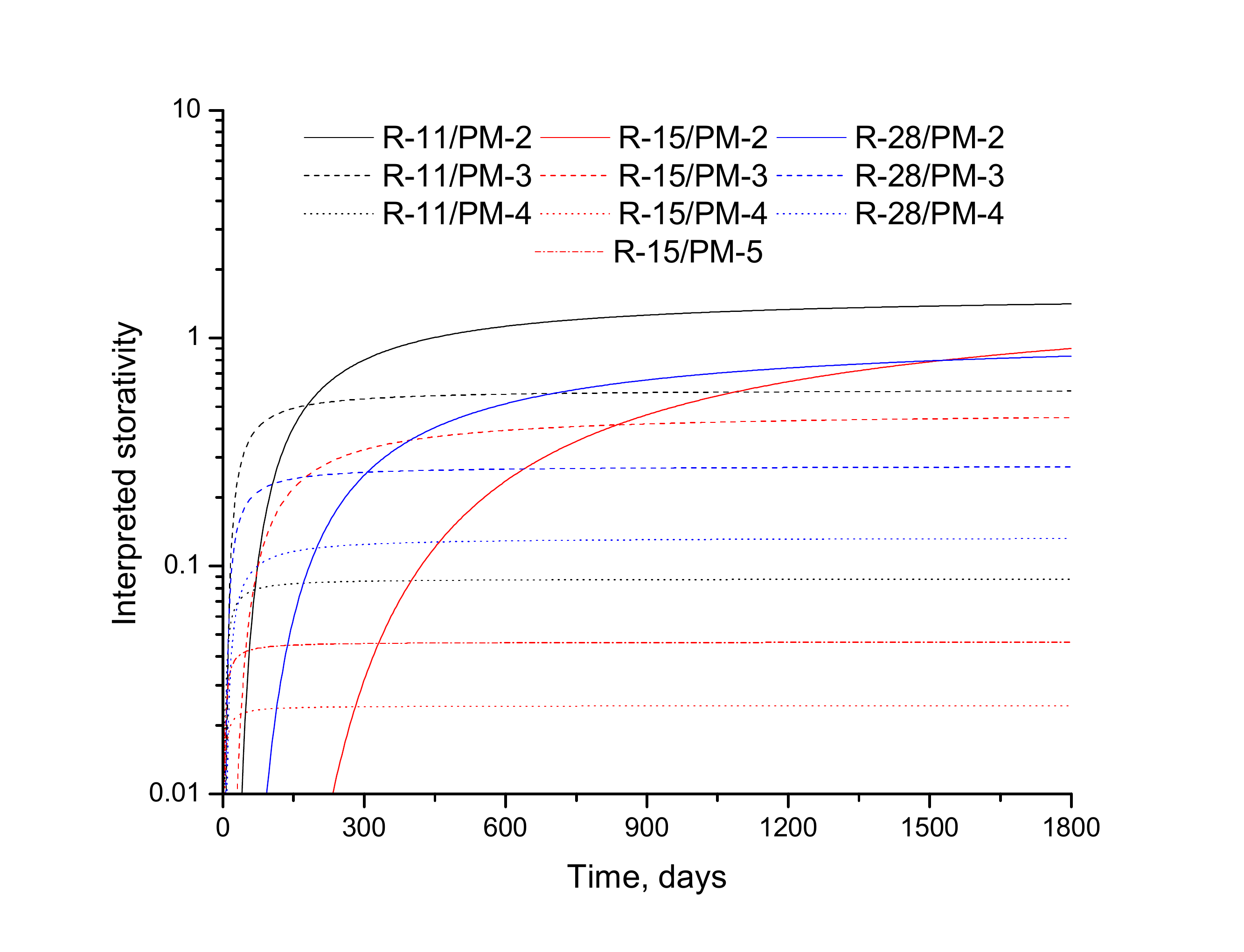}
		\caption{Estimated storativity functions for the exponential case.}
		\label{fig:all-stors}
	\end{center}
\end{figure}

Figures~\ref{fig:exp-calibs} presents the calibrated drawdowns from the water-supply wells for monitoring wells R-11, R-15, and R-28 for the exponential and constant cases. It is apparent that the exponential case reduces the mismatch in drawdowns for all three wells, with the most significant improvements in R-15, the well with the worst match for the constant case. 

Figure~\ref{fig:all-trans} presents the estimated transmissivity functions (equation~(\ref{eq:exp_T})) for R-11, R-15, and R-28. The functions are plotted up to around five years to include parameter values used in the model runs. It is apparent that for all three monitoring wells, $\hat{T}(t)$ converges towards a single value, $\hat{T}_{eff} = 10^{3.07}$~m$^2$/d (1170~m$^2$/d), as constrained by the inversion. The interpreted transmissivity for the constant case ($\hat{T} = 10^{3.27}$~m$^2$/d (1860~m$^2$/d)) is indicated in the figure. It is apparent that this value is fitted to an average value of $\hat{T}(t)$ for the exponential case. This indicates that an overestimate of $T_{eff}$ will be obtained using constant parameters. 

Figure~\ref{fig:all-stors} presents the estimated storativity functions (equation~(\ref{eq:exp_S})). It is apparent that in general the storativity functions converge quickly to distinct values in accordance with previous research \citep{Wu05,Straface07}, providing indications of inter-well connectivity. Physically unrealistic values of storativity are allowed as $\hat{S}_a$ is recognized as a flow connectivity indicator, and does not represent aquifer storativity in an effective or equivalent sense. 


Table~\ref{tab:pars2} presents the estimated parameters associated with the transmissivity and storativity functions plotted in Figures~\ref{fig:all-trans} and \ref{fig:all-stors} for the exponential case and the parameters for the constant case. As constrained in the inversion, all transmissivities converge to a single value for both the exponential and constant cases. For the exponential case, this value can be considered a first estimate of $T_{eff}$. The larger value obtained for the constant case indicates that the calibration has fitted the parameter within the early time variability, thereby overestimating $T_{eff}$. 

Values of $\hat{S}_a$ indicate the level of connectivity between the monitoring and pumping well. Large/small values of $\hat{S}_a$ indicate a region of relatively low/high inter-well transmissivity. It is apparent from Figure~\ref{fig:all-stors} and Table~\ref{tab:pars2} that the trends in $\hat{S}_a$ can be grouped by the associated pumping well. For instance, convergent values of $\hat{S}_a$ decrease (inter-well connectivity increases) from PM-2 to PM-3 to PM-4 and PM-5. In general, similar trends are apparent for $\hat{S}$ in the constant case as well. However, in the constant case, values for PM-2 are farther from physically realistic values of storativity. 

A decomposition of the pressure influences from the pumping wells at the monitoring wells also resulted from this research. These results are similar to the decomposition analysis of this dataset presented in \cite{Harp10} and therefore are not presented here. For instance, the same pumping wells are identified to influence drawdown at the monitoring wells and a lack of a linear temporal trend is identified for R-15\ in both cases. 

\begin{table}
	\begin{center}
	\begin{tabular}{c|c||c|c|c||c|c|c||c|c}
		Monitoring&Pumping&$\hat{T}$&$\hat{T}_{eff}$&$c_T$&$\hat{S}$&$\hat{S}_a$&$c_S$&\multicolumn{2}{c}{$m$[m/a]}\\
		Well&Well&Const.&Exp.&Exp.&Const.&Exp.&Exp.&Const&Exp.\\
		\hline
		\multirow{3}{*}{R-11}&PM-2&\multirow{10}{*}{3.27}&\multirow{10}{*}{3.07}&168.7&2.19&1.58&-203.6&\multirow{3}{*}{0.06}&\multirow{3}{*}{0.03}\\
		&PM-3&&&97.4&0.51&0.60&-28.7&&\\
		&PM-4&&&94.5&0.09&0.09&-7.5&&\\
		\cline{1-2} \cline{5-10}
		\multirow{4}{*}{R-15}&PM-2&&&278.1&4.96&1.76&-1205.4&\multirow{4}{*}{0}&\multirow{4}{*}{0}\\
		&PM-3&&&151.6&0.04&0.48&-116.9&&\\
		&PM-4&&&4.0&0.02&0.02&-3.19&&\\
		&PM-5&&&7.8&0.08&0.05&-4.7&&\\
		\cline{1-2} \cline{5-10}
		\multirow{3}{*}{R-28}&PM-2&&&287.9&3.82&1.06&-433.2&\multirow{3}{*}{0.04}&\multirow{3}{*}{0.02}\\
		&PM-3&&&33.3&0.19&0.27&-19.4&&\\
		&PM-4&&&29.4&0.08&0.13&-21.3&&\\
	\end{tabular}
	\caption{Parameter estimates from calibrations using exponential functions (Exp.) and constant (Const.) parameters. Transmissivity parameters are presented in units of $\log(m^2/d)$. $m$ is the linear temporal trend parameter.}
	\label{tab:pars2}
	\end{center}
\end{table}

\section{Conclusions}	

This paper demonstrates an approach to obtain late-time aquifer property inferences consistent with the Cooper-Jacob method from transient datasets collected in heterogeneous aquifers. Such datasets are commonly available from municipal water-supply networks. The utilization of these existing datasets eliminates the expense and coordination necessary to perform dedicated pumping tests at a site. The methodology is motivated by analytical investigations by \cite{Dagan82}, numerical experiments by \cite{Wu05}, and analysis of field-collected hydrographs by \cite{Straface07}. The hydrogeologic inferences are evaluated based on a large body of research into the meaning of late-time aquifer property inferences \citep{Butler90,Neuman90,Meier98,Sanchez99,Neuman03c,Wu05,Knudby06}.

Utilizing this approach on a dataset from the LANL site has indicated that adequate water-level calibrations can be achieved within the constraints of the inversion: a single value of $\hat{T}_{eff}$ is applied to all pumping/monitoring well pairs; $\hat{T}(t)$ decreases towards a constant value; $\hat{S}_a$ is allowed to take distinct values and is allowed to increase or decrease towards convergent values. $\hat{T}_{eff}$ provides an initial estimate of the effective transmissivity at the support scale characterized by the distances between the pumping and observation wells \citep{Neuman90,Neuman03c}. In accordance with \cite{Meier98}, \cite{Sanchez99}, and \cite{Knudby06}, $\hat{S}_a$ is recognized as an indicator of inter-well connectivity, indicating the degree in which pumping and monitoring well pairs are hydraulically connected.


%
%
%
%
%
%

%
%
%
%

\paragraph{Acknowledgments}
The research was funded through various projects supported by the Environmental Programs Division at the Los Alamos National Laboratory. The authors are thankful for the valuable suggestions and comments provided by Kay Birdsell on draft versions of this paper. The authors are also grateful for constructive comments provided by members of the first authors Ph.D. advisory committee (Bruce Thomson, Gary Weissmann, and John Stormont). 

%
%
\bibliography{bibliography}
\bibliographystyle{agufull04}

\end{document}